\documentclass[aps,prb,twocolumn,superscriptaddress,showpacs]{revtex4-1}

\usepackage{graphicx}
\usepackage{calc}
\usepackage{bm}
\usepackage{color}
\usepackage{textcomp}

\begin{document}

\title{Josephson current along the domain walls in Weyl fully compensated ferrimagnet Ti$_2$MnAl.}

\author{V.D.~Esin}
\author{N.N. Orlova}
\author{A.R. Banschikova}
\author{Yu.S.~Barash}
\author{D.N. Borisenko}
\author{N.N.~Kolesnikov}
\affiliation{Institute of Solid State Physics of the Russian Academy of Sciences, Chernogolovka, Moscow District, 2 Academician Ossipyan str., 142432 Russia}
\author{E.V.~Deviatov}
\affiliation{Institute of Solid State Physics of the Russian Academy of Sciences, Chernogolovka, Moscow District, 2 Academician Ossipyan str., 142432 Russia}
\affiliation{V.L. Ginzburg Research Centre for High-Temperature Superconductivity and Quantum Materials, P.N. Lebedev Physical Institute of RAS, Moscow 119991, Russia}

\date{\today}

\begin{abstract}
We experimentally investigate Josephson current in planar Nb-Ti$_2$MnAl-Nb proximity devices, which are formed as junctions between superconducting niobium leads and  single crystal flakes of the Weyl fully compensated ferrimagnet Ti$_2$MnAl. We demonstrate  well developed Josephson effect for 2~$\mu$m distance between the niobium contacts on the Ti$_2$MnAl surface. The measured critical current dependence on the magnetic field $I_c(B)$ shows strong anisotropy for two magnetic field orientations in respect to the Nb-Ti$_2$MnAl-Nb junction plane.  For the parallel magnetic field, one can see standard Fraunhofer-like $I_c(B)$ suppression pattern. In contrast, $I_c(B)$  is qualitatively different in normal  magnetic field, indicating non-trivial distribution of the Josephson current within the junction plane. From this anisotropy, long-range supercurrent and fast aperiodic fluctuations in $I_c(B)$ we conclude that Josephson current is  transferred by the  chiral surface states, which emerge along the domain walls in topological compensated ferrimagnet Ti$_2$MnAl. The conclusion on the domain walls is supported by magnetization data for small Ti$_2$MnAl flakes.
\end{abstract}


\maketitle

\section{Introduction}

Recent interest to Weyl semimetals (WSM)  is mostly associated with Fermi arc topological surface states,  which connects the projections of Weyl nodes on the surface Brillouin zone~\cite{armitage}. Because of the spin-momentum locking~\cite{spin-burkov},  Weyl semimetals are considered as attractive materials for spin investigations. For example, spin-sensitive angle-resolved photoemission spectroscopy (ARPES) measurements confirm nearly full spin polarization of Fermi arcs~\cite{lv2015,xu16,PhysRevB.100.195134}.

By definition, Weyl semimetals require breaking of the time reversal or inversion symmetries~\cite{armitage}. Most of experimentally investigated WSMs, were non-magnetic non-centrosymmetric crystals~\cite{armitage} while there are a few ferromagnetic Weyl semimetals~\cite{mag1,mag2,mag3,mag4,kagome,kagome1,cosns}. On the other hand,  the topological surface  states of antiferromagnetically (AFM) ordered  materials are regarded as perspective~\cite{AFMspintronics} due to high charge-current to spin-current conversion efﬁciency, electron mobility and efﬁcient spin ﬁltering. Experimentally, the only confirmed realization of the AFM Weyl semimetal~\cite{AFMspintronics} is the family of noncollinear antiferromagnets like Mn$_3$Ge~\cite{mn3ge,mn3ge1,mn3ge2}.  

In proximity with a superconductor, topological surface (or edge) states are able to carry supercurrents over extremely large  distances~\cite{topojj1,inwte1,inwte2,topojj3,incosns,topojj5,rfli}, while the coherence length is much smaller for the bulk carriers. 
Difference in coherence lengths between surface and bulk carriers lead, e.g.,  to the squid-like critical current suppression pattern~\cite{yakoby,kowen,inwte1} and/or to the fractional a.c. Josephson effect~\cite{inwte2,rfpan,rfsnyder}. Thus, Josephson effect allows to reveal charge transport through the surface states even in materials with gapless bulk spectrum. 

Proximized AFM topological materials, in general, could be  highly advantageous for developing superconducting spintronic logic circuits: they produce no stray ﬁelds, which makes them suitable for device miniaturization. As an example, the Josephson spin-valve has recently been realized in  Nb-Mn$_3$Ge-Nb junctions containing a single interlayer of the topological chiral antiferromagnet Mn$_3$Ge~\cite{jsv5a,jsv5}. 

Because of zero net magnetization, the inverse Heusler compound Ti$_2$MnAl is often called as half-metallic antiferromagnet while symmetry requires its formal classification as half-metallic fully compensated ferrimagnet~\cite{timnal_bulk,timnal_bulk1,timnal_bulk2}. Contrary to conventional antiferromagnets, the compensation of the spin magnetic moments stems from different magnetic sublattices in Ti$_2$MnAl, since Mn and Ti atoms have antiparallel spin magnetic moments. Thus, the microscopic magnetization at finite temperatures is not zero as in conventional antiferromagnets~\cite{ferri-antiferro}. In view of the spin splitting, some connection can be seen between the fully compensated ferrimagnets and modern altermagnetic materials~\cite{alter_ferri_symmetry}: in comparison to the noncollinear antiferromagnets like Mn$_3$Ge, bands in compensated ferrimagnet Ti$_2$MnAl are spin split, which allows the existence of Weyl points even without spin-orbit coupling~\cite{timnal,timnal1}.  Ti$_2$MnAl is predicted~\cite{timnal,timnal1} as the magnetic WSM: though the net magnetic moment is zero, the Weyl points lead to a large anomalous Nernst  and Hall effects due to the lack of a global time-reversal symmetry~\cite{timnal,timnal1}.  Topological surface states~\cite{timnal,timnal1} and spin textures have been demonstrated in spin-polarized transport for Ti$_2$MnAl~\cite{timnal_esin}.  

In AFM, domain wall spin-triplet superconductivity and supercurrent-induced torques have been predicted for the spin textures coupled to superconductors,  so  the superconductivity predominantly impacts the regions between AFM domains~\cite{alter_domain_super,domain_wall_supercurrent}. One-dimensional chiral edge states emerge along the domain walls, since domains with opposite Neel vector have opposite topological (Chern)  numbers~\cite{magnetic_Weyl_review,bound_AFM,domain_wall_network,Mn3Ge_domain_wall}.  The domain-wall chiral edge states in topological antiferromagnets can transfer supercurrent~\cite{alter_domain_super,domain_wall_supercurrent}, so they are responsible for nonreciprocal superconducting transport, including asymmetric Andreev spectra, Josephson diode effect, and anomalous Fraunhofer interference patterns~\cite{domain_wall_supercurrent}. Thus, it is also reasonable also to investigate Josephson current in long (multidomain) junctions based on the Weyl fully compensated ferrimagnet Ti$_2$MnAl.

Here, we experimentally investigate charge transport in planar Nb-Ti$_2$MnAl-Nb proximity devices, which are formed as junctions between superconducting niobium leads and  single crystal flakes of the Weyl fully compensated ferrimagnet Ti$_2$MnAl. We demonstrate Josephson current  for 2~$\mu$m distance between the niobium contacts on the Ti$_2$MnAl surface. The measured critical current dependence on the magnetic field $I_c(B)$ is qualitatively different for two magnetic field orientations in respect to the Nb-Ti$_2$MnAl-Nb junction plane. From this observation, fast aperiodic fluctuations in $I_c(B)$ and magnetization data we conclude, that the magnetic domains and transport along the domain walls are  important in planar Nb-Ti$_2$MnAl-Nb junctions.

\section{Samples and characterization}

Ti$_2$MnAl was obtained as a bulk ingot by levitation melting in high-frequency (60-70 kHz) induction furnace. A mixture of Mn and Al powders was placed into the cylindrical titanium capsule and melted in a suspended condition for 20 minutes in argon medium at 0.2 MPa pressure and at 2080 K temperature. After switching the heater off the resulting globule of the melt was dropped to a cooled copper crystallizer, where it was quenched at 278 K. The ingot cleaved mechanically to obtain Ti$_2$MnAl single crystal flakes, which are mostly characterized by well-developed flat surface. 

\begin{figure}
\includegraphics[width=\columnwidth]{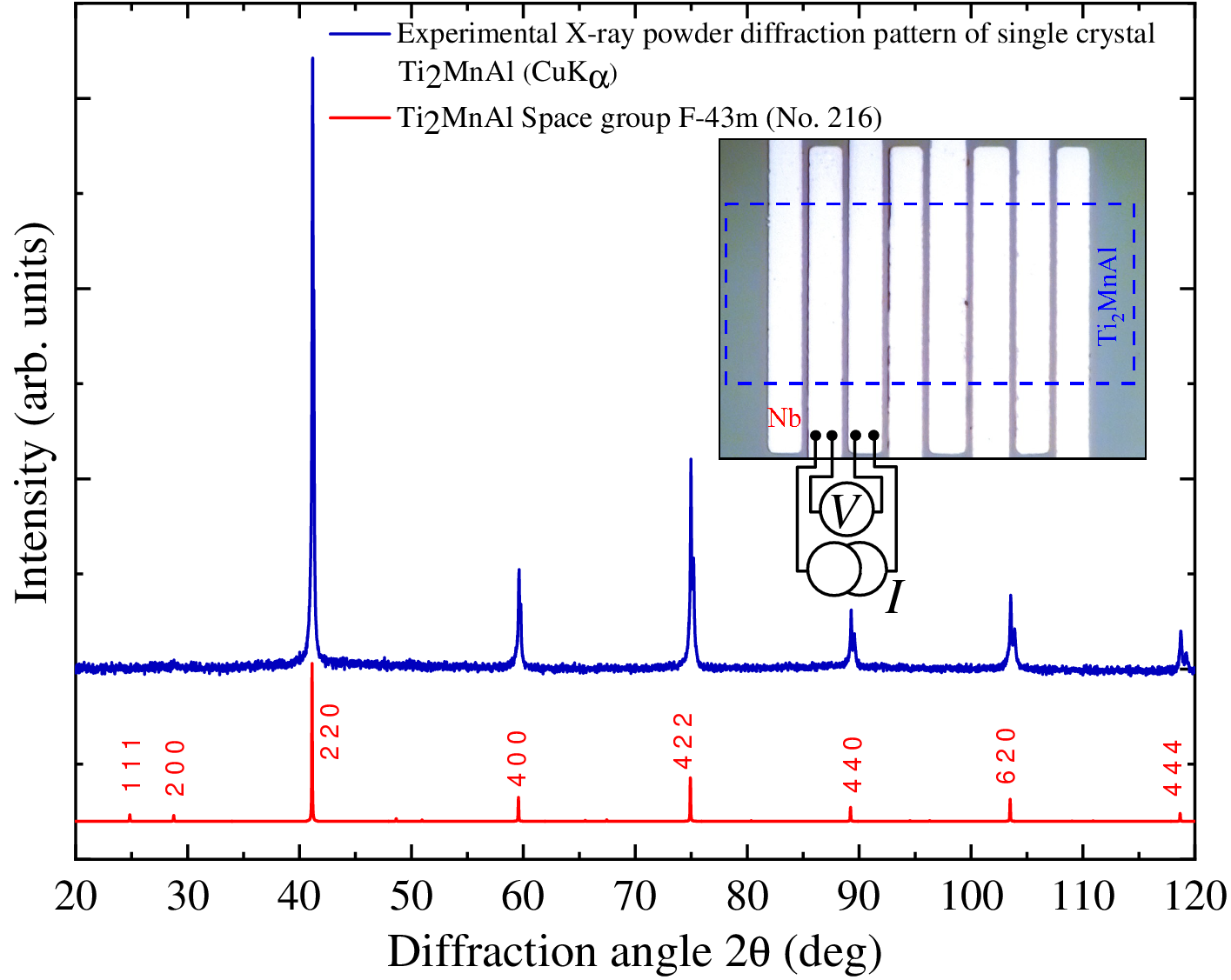}
\caption{(Color online) The X-ray powder diffraction  pattern (Oxford diffraction Gemini-A, CuK$_{\alpha}$ radiation), which is obtained for the crushed Ti$_2$MnAl single crystal. The single-phase Ti$_2$MnAl is confirmed with the space group $F$-$43m$ (No. 216). Inset: sample design for the Josephson current measurements. Ti$_2$MnAl single-crystal flake is placed on the pre-defined Nb leads to form 2~$\mu$m long Nb-Ti$_2$MnAl-Nb junctions.  Differential resistance is measured between two neighbor Nb leads in a standard four-point technique.}
\label{fig1}
\end{figure}

The Ti$_2$MnAl composition is verified by energy-dispersive X-ray spectroscopy. The single-phase Ti$_2$MnAl is confirmed by X-ray powder diffraction  pattern (Oxford diffraction Gemini-A, CuK$_{\alpha}$ radiation) with the space group $F$-$43m$ (No. 216), see Fig.~\ref{fig1}. 

However, despite the correct stoichiometry,   Ti$_2$MnAl single crystal flakes should be firstly characterized in magnetization measurements. For example, the policrystalline Ti$_2$MnAl films usually demonstrate  non-compensated  ferrimagnetic properties with small coercivity and finite magnetization, depending on the film preparation~\cite{timnal_exp,timnal_exp1,timnal_exp2}. Also, the disorder may lead to the broken compensation even in the bulk single-crystal samples~\cite{timnal_bulk3}. 

The magnetic state of Ti$_2$MnAl single crystal flakes is investigated by Lake Shore Cryotronics 8604 VSM magnetometer, equipped with nitrogen flow cryostat. The flake is mounted to the sample holder by low temperature grease. It was verified, that without a sample, the sample holder with corresponding amount of grease shows fully isotropic and strictly linear small diamagnetic response, much below of the measured Ti$_2$MnAl magnetization value.  We investigate sample magnetization by standard method of the magnetic field gradual sweeping between two opposite field values to obtain $M(H)$ magnetization loops at different temperatures and for different angles $\alpha$ between the flake surface and the magnetic field.

\begin{figure}
\includegraphics[width=\columnwidth]{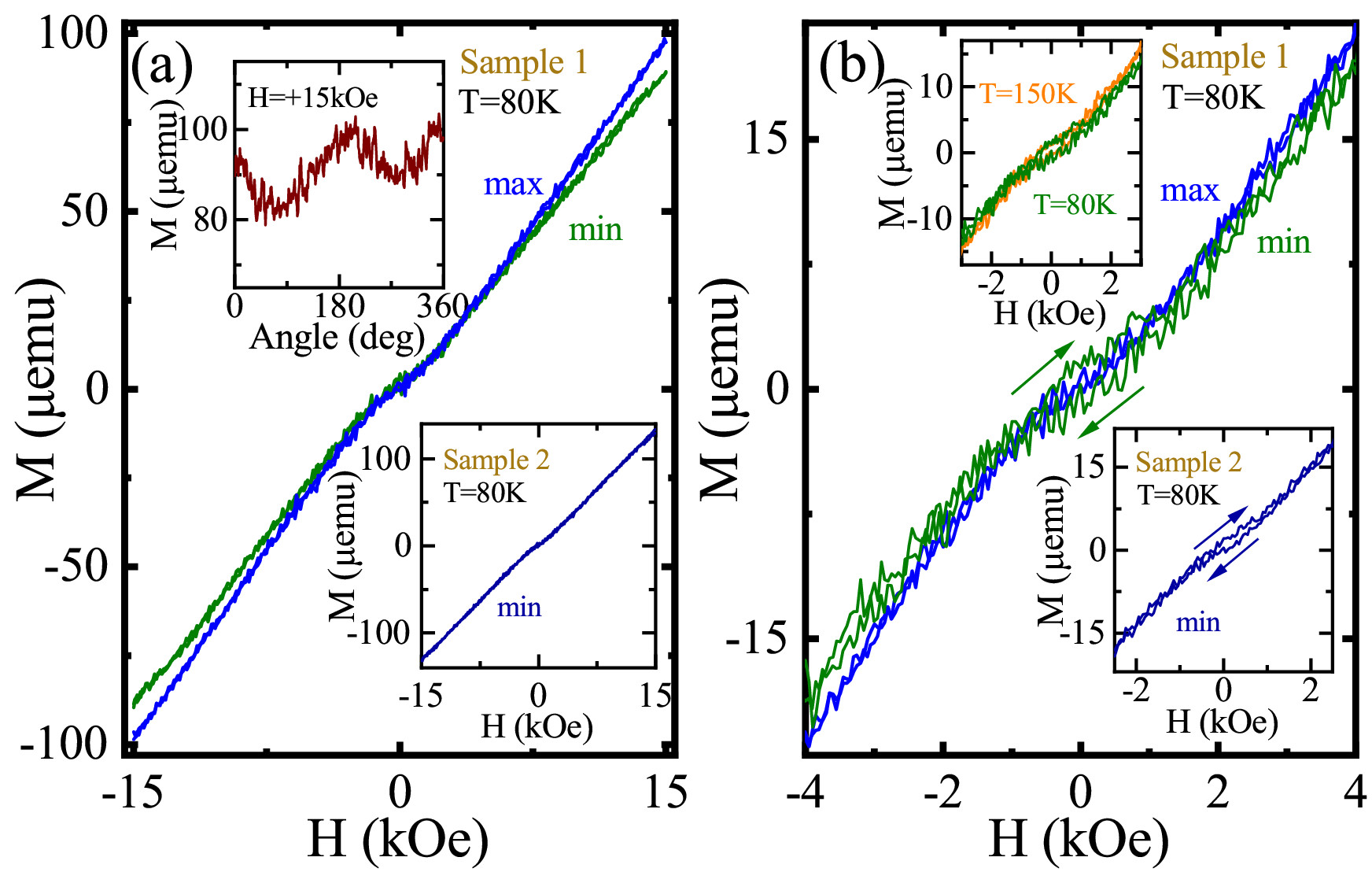}
\caption{(Color online) Magnetization behavior for two Ti$_2$MnAl single-crystal flakes at 80~K temperature, confirming fully compensated ferrimagnetic  state of Ti$_2$MnAl.  
(a) $M(H)$ magnetization loops  for the  minimum (the green curves) and for the  maximum (the blue ones) of the $M(\alpha)$ angle dependence of magnetization for 4.55 mg mass sample.  $M(\alpha)$ is shown in the left inset for 15~kOe magnetic field. For both angles, $M(H)$ is non-linear  in low fields. The right inset shows similar non-linear $M(H)$ magnetization for the bigger,  5.81 mg flake. 
(b) Low-field ($\pm 1$~kOe) inverted hysteresis with negative remanence for the $M(\alpha)$  minimum (the green curves) for 4.55 mg mass sample. In contrast, there is no hysteresis for the $M(\alpha)$  maximum (the blue ones). The hysteresis is not sensitive to temperature below 150~K, as depicted in the left inset.  The right inset shows similar inverted $M(H)$ hysteresis for the  5.81 mg flake. The inverted low-field hysteresis and negative remanence reveal contribution of domain walls between the domains of fully compensated ferrimagnetic state~\cite{negative_remanence_antiferro}.   
}
\label{fig2}
\end{figure} 

Fig.~\ref{fig2} (a) shows $M(H)$ magnetization loops for the 4.55 mg Ti$_2$MnAl single-crystal flake at 80~K temperature. The curves are presented for two angles $\alpha$, which correspond to the  minimum (the green ones) and for the  maximum (the blue ones) of the $M(\alpha)$ angle dependence of magnetization, $M(\alpha)$ is depicted in  the left inset. For both angles, $M(H)$ is non-linear, especially in low fields. The right inset shows similar non-linear $M(H)$ magnetization for the bigger,  5.81 mg Ti$_2$MnAl flake.  

The low-field non-linear region is shown in Fig.~\ref{fig2} (b) for 4.55 mg flake.  For the $M(\alpha)$  minimum (the green curves), we observe inverted (clockwise) hysteresis in $M(H)$ magnetization within $\pm 1$~kOe, while the curves coincide for the $M(\alpha)$ maximum (the blue ones).  The hysteresis is not sensitive to temperature below 150~K, as depicted in the left inset to Fig.~\ref{fig2} (b). The low-field hysteresis can be also demonstrated for the 5.81 mg Ti$_2$MnAl flake, see the right inset. 

This behavior well corresponds to the expected one for  antiferromagnets or  fully compensated ferrimagnets: in the present field range, we do not reach the N\'eel vector reorientation field, so one should expect slightly non-linear high-field branches with positive slope~\cite{AFM_book}. The inverted low-field hysteresis and negative remanence similar to ones in Fig.~\ref{fig2} (b) are known  for AFMs due to non-equilibrium population of domain walls which exhibit a magnetic moment between the domains of an ordered antiferromagnetic state~\cite{negative_remanence_antiferro}. In contrast, the hysteresis is normal (counter-clockwise with positive remanence)  for, e.g., weak ferromagnetism of antiferromagnets~\cite{weak_ferro,weak_ferro1} or the spin-orbit-induced bulk spin polarization in  altermagnets~\cite{alter_ferro,spin_ferro_soc,orlova_MnTe1,orlova_MnTe2}. 

Thus, magnetization measurements  not only confirm  full ferrimagnetic compensation for our Ti$_2$MnAl single crystal flakes but also reveal contribution of domain walls between the domains of fully compensated ferrimagnetic state.

\section{Experimental results: Josephson current}

The topological surface states are able  to carry the Josephson current  even for several micrometers distances~\cite{topojj1,inwte1,inwte2,topojj3,incosns,topojj5,rfli}. 

Inset to Fig.~\ref{fig1} shows schematically sample design for the Josephson current measurements. The 10~$\mu$m wide Nb leads are separated by 2~$\mu$m intervals, they are formed by lift-off technique after sputtering of 100~nm thick Nb on a standard Si/SiO$_2$ substrate. We select relatively thick (above 1~$\mu$m) Ti$_2$MnAl single crystal flakes with $\approx$100~$\mu$m lateral size with mostly plane-parallel, defect-free surface. The flake  is placed on the Nb contact leads. After initial single-shot pressing by another oxidized silicon substrate,  the flake is firmly connected to the leads. This procedure provides transparent Nb-Ti$_2$MnAl junctions, stable in subsequent cooling cycles, which has been verified for a wide range of materials~\cite{inwte1,inwte2,incosns,nb_wte2,aunite,cosi}. As an additional advantage, the Nb-Ti$_2$MnAl-Nb junctions are protected from any contamination by the Si/SiO$_2$ substrate and Ti$_2$MnAl bulk. 

The Nb-Ti$_2$MnAl-Nb junction resistance is measured in a standard four-point technique~\cite{aunite}, as schematically presented in the inset to  Fig.~\ref{fig1}. All the wire resistances are excluded, which is necessary for low-impedance samples. To obtain differential $dV/dI(I)$ characteristics, dc current is additionally modulated by a low (100~nA) ac component, so the lock-in detected ac voltage is proportional to the differential resistance $\sim dV/dI$. The signal is confirmed to be independent of the modulation frequency within 100 Hz -- 10kHz range, which is defined by the applied filters.  The measurements below are performed in a standard liquid helium cryostat for two orientations of the magnetic field in respect to the Nb-Ti$_2$MnAl-Nb junctions plane. The temperatures below 1.2~K are obtained in a dilution refrigerator.

\begin{figure}
\includegraphics[width=\columnwidth]{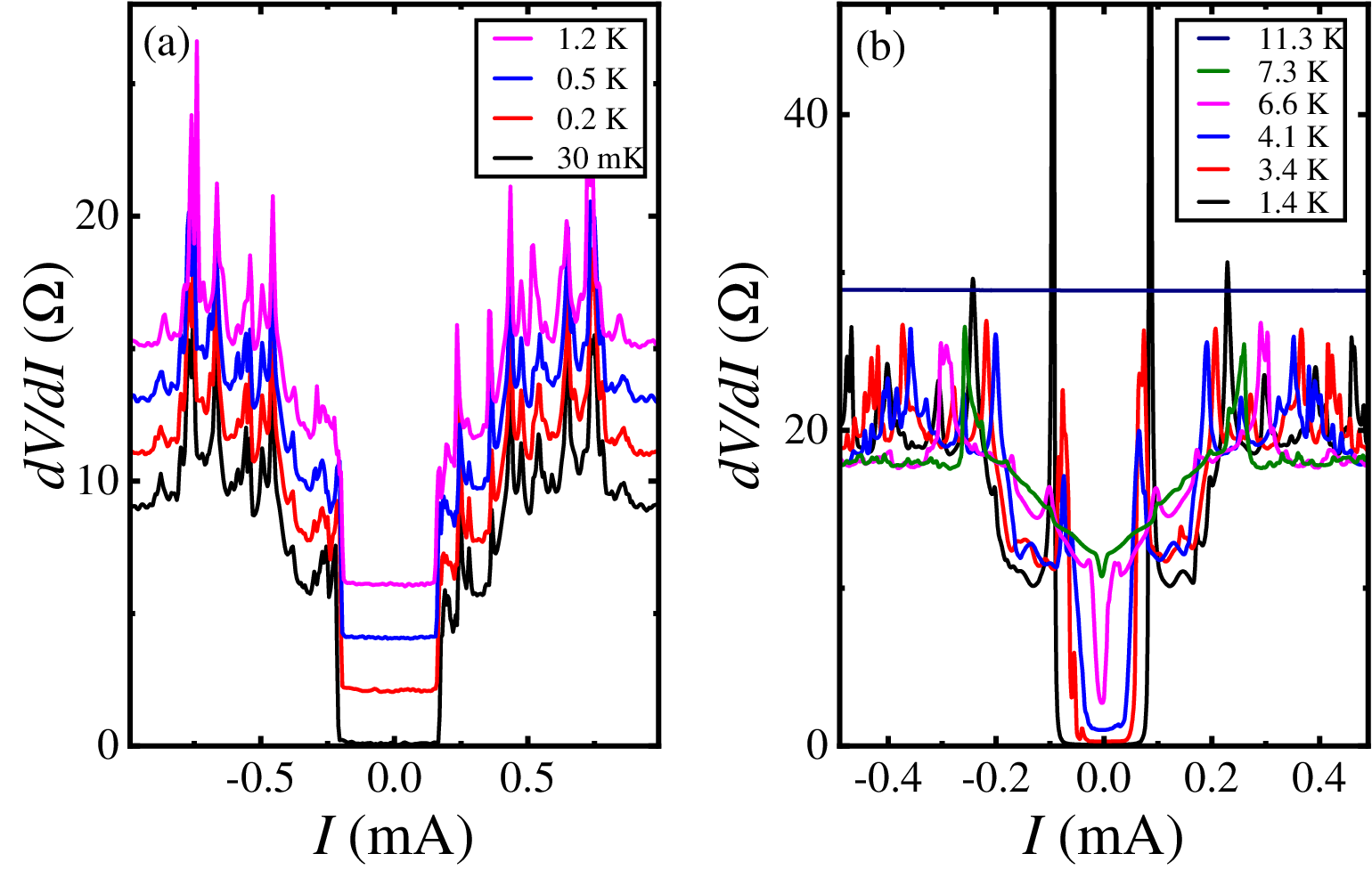}
\caption{(Color online) Josephson $dV/dI(I)$ curves at different temperatures for the 2~$\mu$m long Nb-Ti$_2$MnAl-Nb junction. (a) Both the zero-resistance region and the resonant peaks' positions well coincide at 30~mK -- 1.2~K. The curves are shifted vertically for clarity. (b) Temperature dependence above 1.2~K. The zero-resistance region is suppressed  above 3.4~K, the  $dV/dI(I)$ curve is of standard Andreev shape~\cite{tinkham} at 7.3~K, there is no any signs of superconductivity at 11.3~K, i.e. above the Nb critical temperature. The curves are obtained in zero magnetic field.}
\label{fig3}
\end{figure}

Fig.~\ref{fig3} (a)  demonstrates Josephson $dV/dI(I)$ curves for  the Nb-Ti$_2$MnAl-Nb planar junction  at several temperatures in the 30~mK -- 1.2~K range. The zero-resistance state appears below some current $I_c$. The absolute values of  $I_c$ are different for resistive-superconducting and superconducting-resisitve transitions, so the zero-resistance region is  slightly asymmetric due to the current sweep direction in Fig.~\ref{fig3} (a). The $I_c$ values are much below the critical current of the Nb leads, which we can estimate for the present geometry as about 1.5 -- 2~mA from the independent measurements. The $dV/dI(I)$ resistive branches demonstrate a set of resonances, which usually belong to multiple Andreev reflection~\cite{tinkham}.

The $dV/dI(I)$ curves are  independent of temperature below 1.2~K in Fig.~\ref{fig3} (a), because of high critical temperature for niobium (9~K for the present Nb film). Both the zero-resistance region and the resonant peaks' positions well coincide at 30~mK -- 1.2~K, demonstrating also the stability of our Nb-Ti$_2$MnAl-Nb devices.  Fig.~\ref{fig3} (b) shows temperature dependence of the  $dV/dI(I)$ curves for higher temperatures. The zero-resistance region is suppressed  above 3.4~K,  the  $dV/dI(I)$ curve is of standard Andreev shape~\cite{tinkham} at 7.3~K, the superconductivity is fully suppressed  at 11.3~K, i.e. when the temperature exceeds the Nb critical value.

\begin{figure}
\includegraphics[width=\columnwidth]{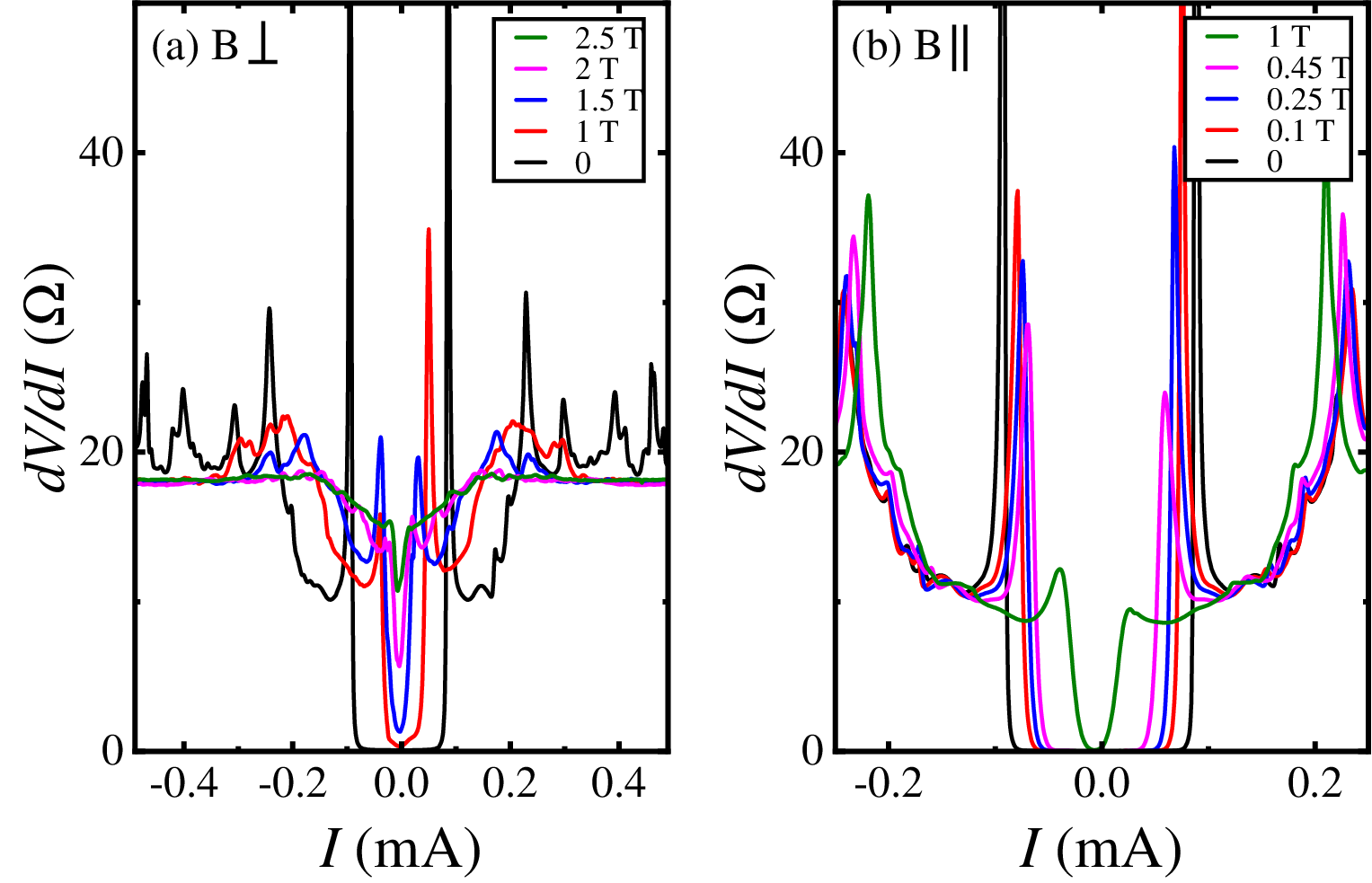}
\caption{(Color online) Magnetic field suppression of the Josephson effect at 1.4~K for two field orientations. Magnetic field is directed normally to the Nb-Ti$_2$MnAl-Nb plane in (a), while it is within the Nb-Ti$_2$MnAl-Nb plane in (b). There is no visible anisotropy in the  $dV/dI(I)$ response  to the  magnetic field, the zero-resistance region is suppressed  at the same 1~T field for both field orientations.   In (b), the enlarged region of low currents is shown to highlight the slow evolution of the experimental $dV/dI(I)$ curves in low magnetic fields.  
}
\label{fig4}
\end{figure}

$dV/dI(I)$ curves can be also affected by magnetic field, as it is demonstrated in Fig.~\ref{fig4} for two field orientations. Magnetic field is directed normally to the Nb-Ti$_2$MnAl-Nb plane (i.e. to the Ti$_2$MnAl flake surface) in Fig.~\ref{fig4} (a), while the field is parallel to the plane in Fig.~\ref{fig4} (b). The zero-resistance region is suppressed  at the same 1~T field for both field orientations, so there is no visible anisotropy in the  $dV/dI(I)$ response  to the  magnetic field. 

As a result,  Nb-Ti$_2$MnAl-Nb junction demonstrates Josephson current for large, 2~$\mu$m distance between the Nb leads on the surface of Weyl fully compensated ferrimagnet Ti$_2$MnAl. 

\begin{figure}
\includegraphics[width=\columnwidth]{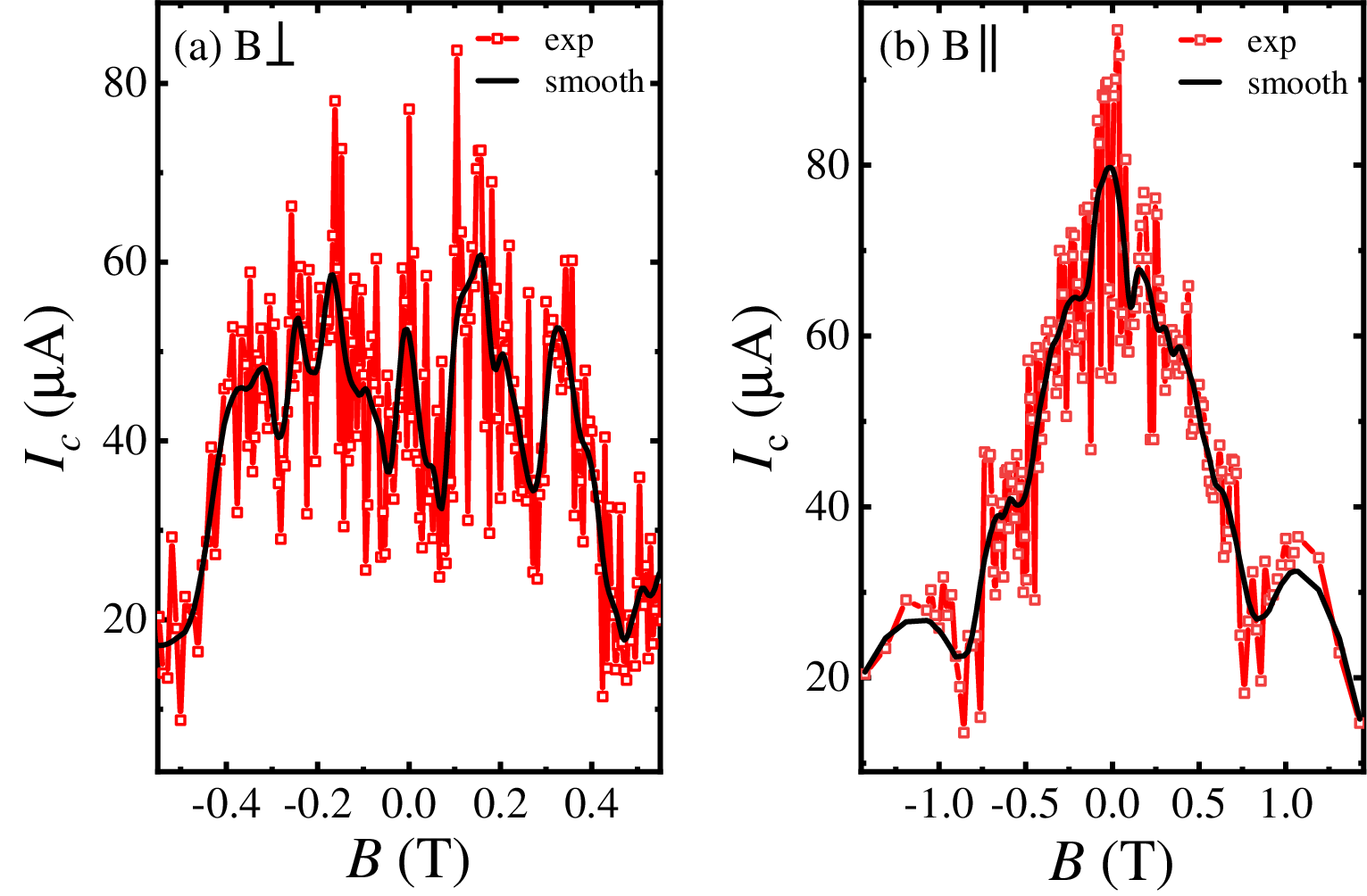}
\caption{(Color online) Critical current dependence on the magnetic field $I_c(B)$ for (a) normal to the Nb-Ti$_2$MnAl-Nb plane magnetic field and (b) parallel one. The original data are shown by red color, they demonstrate fast aperiodic fluctuations~\cite{aperiodicsfs1,aperiodicsfs2,infgt} despite the obtained $I_c$ values are perfectly stable at fixed magnetic field. The black curves show the smoothed data: Fraunhofer pattern~\cite{tinkham,cuevas} can be clearly seen in (b) while $I_c(B)$ is even qualitatively different in (a), indicating non-trivial distribution of the Josephson current within the junction plane~\cite{yakoby,kowen,inwte1}. The data are obtained at T=1.4~K.}
\label{fig5}
\end{figure}

Due to the slow evolution of the experimental $dV/dI(I)$ curves, see Fig.~\ref{fig4} (b), it is possible to obtain the critical current dependence on the magnetic field $I_c(B)$.  The Josephson critical current $I_c$ is measured for the transition from the superconducting $dV/dI= 0$ state to the resistive $dV/dI > 0$ one.  To obtain $I_c$ with high accuracy we sweep the dc current 10 times from zero to some value well above the critical current $I_c$  at fixed magnetic field $B$. Then, we determine $I_c$ for this field $B$ as an average value of $dV/dI$ breakdown positions. The obtained $I_c(B)$ data are shown in Fig.~\ref{fig5} for two magnetic field orientations to the  Nb-Ti$_2$MnAl-Nb junction plane. 

First of all, $I_c(B)$ shows fast aperiodic fluctuations in Fig.~\ref{fig5} (a) and (b), as depicted by red color. We check, that our procedure gives $I_c$ values, which are perfectly stable at fixed magnetic field. The maximum $I_c$ deviation over 10 curves at given magnetic field is negligible comparing to the observed fluctuations' amplitudes in Fig.~\ref{fig5}. Thus, the fast fluctuations are controlled by the external magnetic field, although they are found to be aperiodic~\cite{aperiodicsfs1,aperiodicsfs2,infgt}.

Secondly, $I_c(B)$  pattern strongly depends on the magnetic field orientation to the  Nb-Ti$_2$MnAl-Nb junction plane. For the parallel magnetic field, one can see standard~\cite{tinkham} Fraunhofer-like $I_c(B)$ suppression pattern~\cite{cuevas} in Fig.~\ref{fig5} (b) even despite the  fluctuations. After smoothing the data (as depicted by the black curve), Fraunhofer pattern can be clearly seen, which is usual for  Josephson junctions due to dephasing effects~\cite{cuevas}. In contrast, $I_c(B)$ suppression pattern is qualitatively different in normal to the Nb-Ti$_2$MnAl-Nb plane magnetic field: $I_c(B)$ is diminishing slowly within $\pm 0.4$~T, with abrupt drop outside this region. The smoothed curve also oscillate in  Fig.~\ref{fig5} (a), resembling squid-like behavior due to the non-trivial distribution of the Josephson current within the junction plane~\cite{yakoby,kowen,inwte1}.

\begin{figure}
\includegraphics[width=\columnwidth]{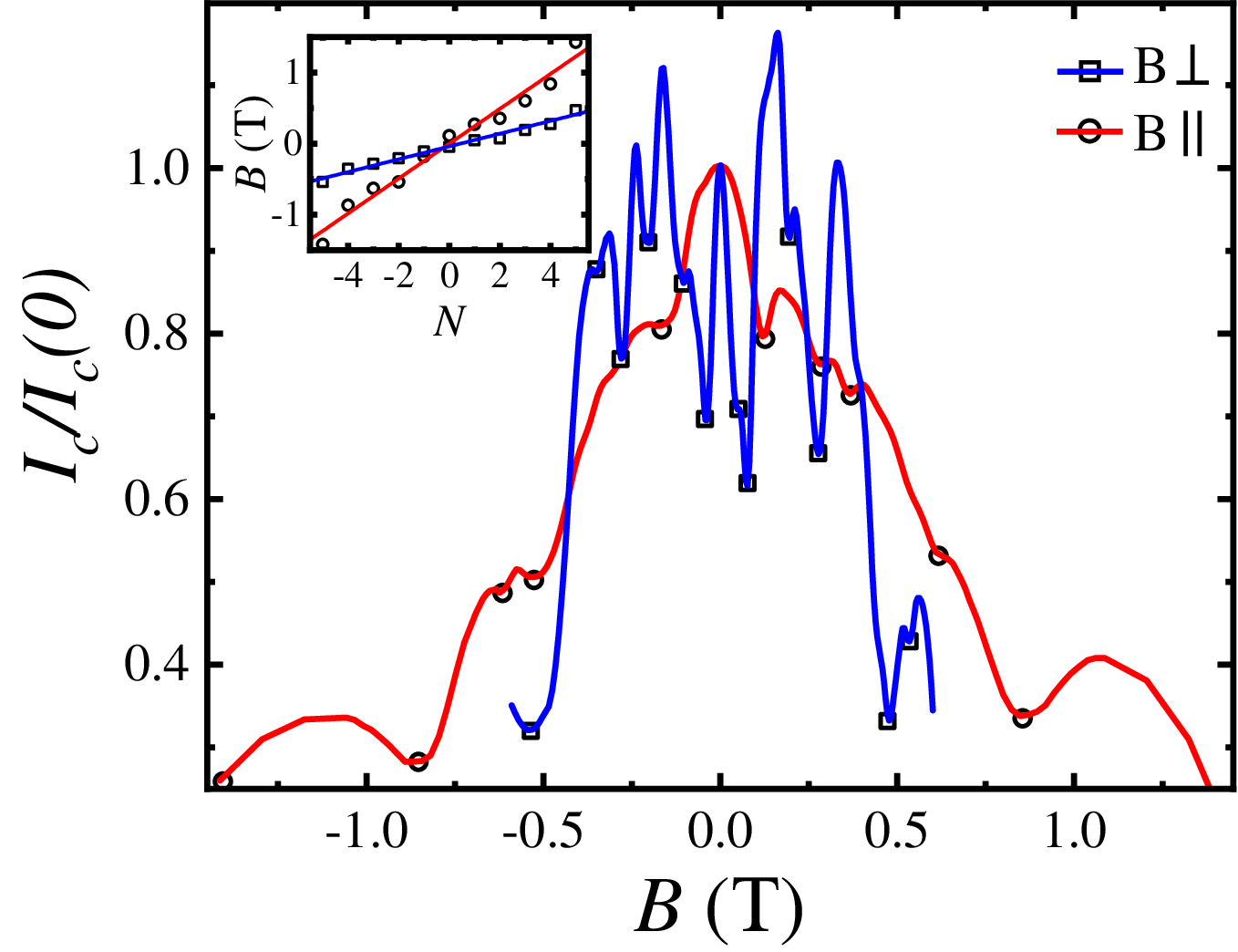}
\caption{(Color online) Comparison of the smoothed $I_c(B)$ data for normal (blue) and parallel (red) magnetic field orientations.
$I_c(B)$ patterns are qualitatively different: : $I_c(B)$ is diminishing slowly within $\pm 0.4$~T, with abrupt drop outside this region in normal magnetic field, in contrast to standard Fraunhofer pattern in the parallel one. For both orientations, there are equidistant shallow $I_c(B)$ oscillations, the minima  are denoted by open symbols. Inset shows the minima positions as function of their numbers, which allows to obtain $\Delta B=0.25$~T period for the parallel field and much smaller $\Delta B=0.09$~T for the normal one.
}
\label{fig6}
\end{figure}

The smoothed $I_c(B)$ data are compared in Fig.~\ref{fig6} for two magnetic field orientations to the  Nb-Ti$_2$MnAl-Nb junction plane. For both orientations, we observe equidistant $I_c(B)$ oscillations. The oscillations are characterized by  high $\Delta B=0.25$~T period for the parallel field and by low $\Delta B=0.09$~T for the perpendicular one, see the inset to Fig.~\ref{fig6}. The effective area can be estimated from these periods, $S\Delta B \sim \Phi_0=h/2e$, which gives $S\approx 0.4\cdot 10^{-10}\mbox{cm}^{2}$  for  $\Delta B=0.25$~T in parallel magnetic field. For the micrometer-length junction, the obtained $S$ is equivalent to the thickness of several nanometers, so the Josephson current is indeed concentrated near the sample surface.

\section{Discussion}

As a result of the experiment, we demonstrate Josephson current on the  surface of fully compensated ferrimagnet Ti$_2$MnAl for 2~$\mu$m separated niobium contacts. The critical current dependence on the magnetic field $I_c(B)$ is qualitatively different for two magnetic field orientations in respect to the Nb-Ti$_2$MnAl-Nb junction plane. For both orientations, $I_c(B)$ shows fast aperiodic fluctuations, which are controlled by magnetic field.

In principle, magnetic field anisotropy could be expected  for the planar Nb-Ti$_2$MnAl-Nb junction geometry. However, the zero-resistance region is suppressed  at the same 1~T field for both field orientations in Fig.~\ref{fig4}, so the geometry of the Nb leads (i.e. vortex penetration effects, etc) seems to be insignificant to explain the qualitatively different $I_c(B)$ patterns in normal and parallel magnetic field orientations~\cite{vortex1,vortex2,vortex3}. 

It is worth noting, that it is difficult to expect the parasite shorting of Nb leads for 2~$\mu$m distances. If the Nb short circuit existed, $I_c$ would be almost independent of the magnetic field within the given range~\cite{jsv5a}, also, strong asymmetry between the critical and reverse currents would be observed~\cite{vortex_induced}, in contrast to the data in Figs.~\ref{fig3} and~\ref{fig4}. Also, niobium is a conventional s-wave superconductor, so the fast aperiodic fluctuations in Fig.~\ref{fig5} can not be expected for 10~$\mu$m wide niobium leads~\cite{vortex3}.  Thus, the observed effects should mainly be associated with the specific properties of the proximized fully compensated ferrimagnet Ti$_2$MnAl surface.

Despite Ti$_2$MnAl is characterized by conductive bulk, one cannot expect the bulk Josephson current for 2~$\mu$m long Nb-Ti$_2$MnAl-Nb  junctions~\cite{kulik-long,dubos}, because of much smaller bulk coherence length $\xi<<L$.   On the other hand, Ti$_2$MnAl is predicted~\cite{timnal,timnal1} as the magnetic Weyl semimetal. Topological surface states can efficiently transfer Josephson current for micrometer distances, because the topological protection can significantly increase the coherence length~\cite{topojj1,inwte1,inwte2,topojj3,incosns,topojj5,rfli}. 

For usual Josephson junctions, one could expect standard Fraunhofer-like $I_c(B)$ suppression pattern~\cite{tinkham,cuevas}, as we indeed observe for the parallel magnetic field, see Fig.~\ref{fig6}.   Qualitatively different  $I_c(B)$ suppression pattern in normal magnetic field should be  connected with non-trivial distribution of the Josephson current within the junction plane: if there are several Josephson channels between the niobium leads, normal  magnetic field induces a phase shift between  them, producing squid-like $I_c(B)$ suppression pattern~\cite{yakoby,kowen}. 

The Weyl topological surface states are two-dimensional, so they can not be directly responsible for the  non-trivial distribution of the Josephson current within the Nb-Ti$_2$MnAl-Nb plane, in contrast to one-dimensional edge states in Refs.~\cite{yakoby,kowen}. On the other hand, the effective area $S\approx 10^{-10}\mbox{cm}^{2}$  for  $\Delta B=0.09$~T in normal magnetic field  is equivalent to the lateral distance of about 10~nm between the conductive channels, which is comparable with the antiferromagnetic domain size~\cite{AFM_domain_review}.  

The statement on domains is also confirmed~\cite{negative_remanence_antiferro} by the inverted low-field hysteresis and negative remanence in Fig.~\ref{fig2} (b). The latter effects are usually connected with non-equilibrium population of domain walls which exhibit a magnetic moment between the domains of an ordered antiferromagnetic state~\cite{negative_remanence_antiferro}.  The domain walls dinamics~\cite{instabilities_ferri} could noticeably contribute to aperiodic variations presented in Fig.~\ref{fig5}, similarly to stochastic variations of  $I_c(B)$ in Refs.~\cite{aperiodicsfs1,aperiodicsfs2,infgt} for SFS junctions with multi domain barriers. From these considerations, one should conclude, that the magnetic domains and transport through the domain walls are  important in planar Nb-Ti$_2$MnAl-Nb junctions.

\begin{figure}
\includegraphics[width=\columnwidth]{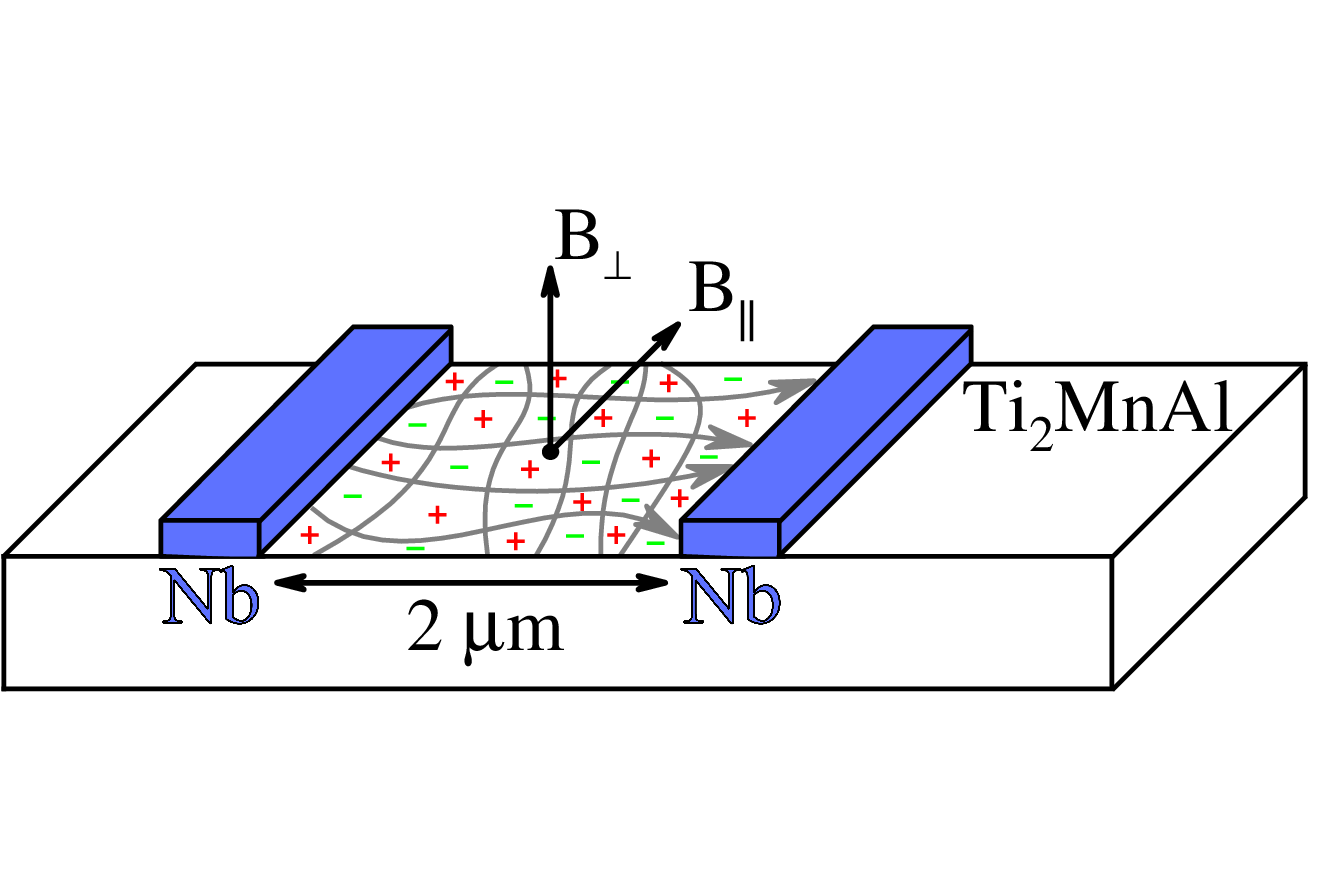}
\caption{(Color online) Schematic diagram of Josephson current transfer by chiral surface states in planar Nb-Ti$_2$MnAl-Nb junction. One-dimensional chiral edge states (arrows) emerge along the domain walls (bold curves), since domains with opposite Neel vector (denoted by + and - simbols) have opposite topological Chern  numbers~\cite{magnetic_Weyl_review,bound_AFM,domain_wall_network,Mn3Ge_domain_wall}.  The domain-wall chiral edge states in topological antiferromagnets can transfer supercurrent~\cite{alter_domain_super,domain_wall_supercurrent}, so multiple weak links connect the superconducting Nb leads (blue rectangles) on the Ti$_2$MnAl surface. In normal magnetic field $B_\perp$, multiple domain walls produce squid-like  oscillations in Figs.~\ref{fig5} (a) and~\ref{fig6}. No phase shift can be expected between the domain walls in parallel field $B_{||}$, so the $I_c(B)$ suppression pattern is of Fraunhofer type in Figs.~\ref{fig5} (b) and~\ref{fig6}. The domain walls dinamics~\cite{instabilities_ferri} could noticeably contribute to aperiodic variations  in Fig.~\ref{fig5}.
}
\label{fig7}
\end{figure}

The  domains with comparatively small sizes are randomly distributed over the surface~\cite{AFM_domain_review}, they are characterized by opposite Neel vector, and, therefore, opposite topological (Chern) numbers, see the schematic diagram in Fig.~\ref{fig7}. One-dimensional chiral edge states emerge along the domain walls~\cite{magnetic_Weyl_review,bound_AFM,domain_wall_network,Mn3Ge_domain_wall}, they can transfer supercurrent~\cite{alter_domain_super,domain_wall_supercurrent} over large distances due to the topological protection. 
When the domain walls are parallel to the current direction, each domain wall introduces additional channel for the Josephson current running along it, see Fig~\ref{fig7}. 

In this case, multiple weak links connect the superconducting niobium leads~\cite{domain_wall_supercurrent}, producing squid-like  oscillations in Figs.~\ref{fig5} (a) and~\ref{fig6} in normal to the  Nb-Ti$_2$MnAl-Nb plane magnetic field.  Since these Josephson channels are within the Nb-Ti$_2$MnAl-Nb plane, no phase shift could be expected in parallel fields, so the $I_c(B)$ suppression pattern is of Fraunhofer type in Fig.~\ref{fig5} (b). Estimation of the effective area from the oscillation period gives reasonable thickness of several nanometers for the current-carrying layer, i.e. thickness of the Weyl surface states for the parallel field configuration.

It is worth noting, that  different orientation of the domain walls (i.e. domain-wall chiral edge states) in respect to the Nb leads  allows to realize both 0- and $\pi$- junctions within the same Nb-Ti$_2$MnAl-Nb junction plane~\cite{pi_1}. In this case, the critical current minimum at zero applied field and its initial increase at finite normal fields in Fig.~\ref{fig6} could reflect the competition between the 0- and $\pi$- contributions~\cite{pi_2}.  

The Josephson effect has recently been observed in the AFM-based Nb-Mn$_3$Ge-Nb junctions~\cite{jsv5,jsv5a}. The authors attributed this observation primarily to a proximity-induced, spin-polarised triplet supercurrent carried trough a single chiral antiferromagnetic interlayer of the topological chiral antiferromagnet Mn$_3$Ge. At the same time, the role of topological surface states and domain walls~\cite{Mn3Ge_domain_wall} was not considered for relatively short Nb-Mn$_3$Ge-Nb junctions~\cite{jsv5a,jsv5}. In contrast, in our experiments the topological states along the domain walls play a central role in establishing the Josephson supercurrent for 2~$\mu$m long Nb-CrSb-Nb junctions.

\section{Conclusion}

As a conclusion, we experimentally investigate Josephson current in planar Nb-Ti$_2$MnAl-Nb proximity devices, which are formed as junctions between superconducting niobium leads and  single crystal flakes of the Weyl fully compensated ferrimagnet Ti$_2$MnAl. We demonstrate  well developed Josephson effect for 2~$\mu$m distance between the  niobium contacts on the Ti$_2$MnAl surface. The measured critical current dependence on the magnetic field $I_c(B)$ shows strong anisotropy for two magnetic field orientations in respect to the Nb-Ti$_2$MnAl-Nb junction plane.  For the parallel magnetic field, one can see standard Fraunhofer-like $I_c(B)$ suppression pattern. In contrast, $I_c(B)$  is qualitatively different in normal  magnetic field, indicating non-trivial distribution of the Josephson current within the junction plane. From this anisotropy, long-range supercurrent and fast aperiodic fluctuations in $I_c(B)$ we conclude that Josephson current is  transferred by the  chiral surface states, which emerge along the domain walls in topological compensated ferrimagnet Ti$_2$MnAl. The conclusion on the domain walls is supported by magnetization data for small Ti$_2$MnAl flakes.

\acknowledgments
We wish to thank S.V.~Simonov for X-ray sample characterization.  We gratefully acknowledge financial support  by the  Russian Science Foundation, project RSF-26-12-00001, https://rscf.ru/project/26-12-00001/. The X-ray powder diffraction pattern was obtained using equipment of the shared facility center of ISSP RAS.

\end{document}